\documentclass[a4paper]{article}

\usepackage{INTERSPEECH2020}

\usepackage{multirow,comment}
\title{Weakly Supervised Training of Hierarchical Attention Networks for Speaker Identification}
\name{Yanpei Shi, Qiang Huang, Thomas Hain}
\address{ Speech and Hearing Research Group\\
          Department of Computer Science, University of Sheffield}
\email{\{YShi30, qiang.huang, t.hain\}@sheffield.ac.uk}

\begin{document}

\maketitle
\begin{abstract}
Identifying multiple speakers without knowing where a speaker's voice is in a recording is a challenging task. 
In this paper, a hierarchical attention network is proposed
to solve a weakly labelled speaker identification problem. 
The use of a hierarchical structure, consisting of a frame-level encoder
and a segment-level encoder, aims to learn
speaker related information locally and globally.
Speech streams are segmented into
fragments.
The frame-level encoder with
attention learns features and highlights the target related frames locally,
and output a fragment based embedding.
The segment-level encoder works with a second attention layer to
emphasize the fragments probably related to target speakers.
The global information is finally collected from
segment-level module to predict speakers via a classifier.
To evaluate the effectiveness of the
proposed approach, artificial datasets based on Switchboard
Cellular part1 (SWBC) and Voxceleb1 are constructed in two
conditions, where speakers' voices are overlapped and not overlapped.
Comparing to two baselines the obtained results show that the
proposed approach can achieve better
performances.
Moreover, further experiments are conducted to
evaluate the impact of utterance segmentation.
The results show that a reasonable segmentation can slightly improve
identification performances.

\end{abstract}
\noindent\textbf{Index Terms}: Weakly Supervised Learning, Speaker Identification, Hierarchical Attention, X-vectors, Attention Mechanism

\vspace*{-1mm}
\section{Introduction}
\vspace*{-1mm}
Speaker identification using deep neural networks becomes an active research area in recent years \cite{variani2014deep,wang2018attention}. 
In traditional supervised speaker identification training, the data used for training needs hand labelling, where the segments and the corresponding speaker labels are manually annotated \cite{karu2018weakly}. It might be expensive to process a large dataset with a large number of speakers using hand annotation \cite{karu2018weakly,jia2019leveraging}.

Instead of hand annotating speaker labels in supervised training, weakly supervised training only relies on the set of speaker labels that occur in the corresponding utterance \cite{zhou2018brief}. This kind of weakly labelled large data collections are available online \cite{karu2018weakly}. Making use of such data collections would be helpful for training with a large amount of data. 

Weakly supervised training has been widely used in speech technology. In \cite{karu2018weakly}, Karu et.al  proposed a DNN based weakly supervised speaker identification training technique. In their work, speaker diarization is firstly applied, and i-vectors are then extracted for each segments. A DNN is trained to predict the set of speaker labels without the true mapping from the i-vectors to the speaker labels. 
In \cite{xu2017unsupervised}, Xu et,at. proposed a DNN based approach for multi-label audio tagging. In their work, an auto-encoder is trained to predict multiple labels using one input utterance. In \cite{xu2018large}, Xu et al. proposed to use a gated convolutional neural network for audio classification. In their work, the model is trained to predict one or more classes from an audio without time stamp labels. 

Except for speech technology, weakly supervised learning has been widely used in other domains.
In \cite{liu2019weakly}, Liu et,al. proposed a weakly supervised transfer learning approach to classify multi-temporal remote-sensing images using one labelled image. In \cite{xu2019weakly}, Xu et,al. proposed a weakly supervised training approach for image semantic segmentation using image-level labels.

In this work, a hierarchical attention network \cite{yang2016hierarchical} based weakly supervised speaker identification approach is proposed. In the training and test data, each utterance contains multiple speakers and only the utterance-level labels are available. Different speakers might occur in different part of the input utterance, and some segments might contain multiple overlapped speakers. The model is trained to predict the set of all of the speaker labels from one input utterance \cite{zhang2007multi,xu2017unsupervised}. The proposed hierarchical attention network contains a frame-level encoder with attention, and a segment-level encoder with attention, which capture speaker information locally and globally \cite{shi2020h}. The frame-level encoder with attention tries to find the important frames within a segment, and the segment-level encoder tries to find the most important parts in the input utterance for speaker identities. Finally, the whole input utterance is compressed into a single vector and input to a DNN classifier. The score vector for each speaker is obtained using a sigmoid function. The proposed hierarchical attention network (HAN) enables the model to highlight and pay attention to the most important parts of input utterance relates the speaker identities.

The rest of the paper is organized as follow: Section \ref{Model Architecture} 
presents the architecture of our approach. 
Section \ref{Experiments} depicts the data and the data construction process, the experimental setup, the baselines to be compared and implementation details.
The results are obtained and shown in Section \ref{Results}, and a conclusion is in Section \ref{Conclusion and Future Work}.

\vspace*{-3mm}
\section{Proposed Model}\label{Model Architecture}
\vspace*{-1.5mm}
\begin{figure}[h]
\centering
\includegraphics[height=10cm,width=7cm]{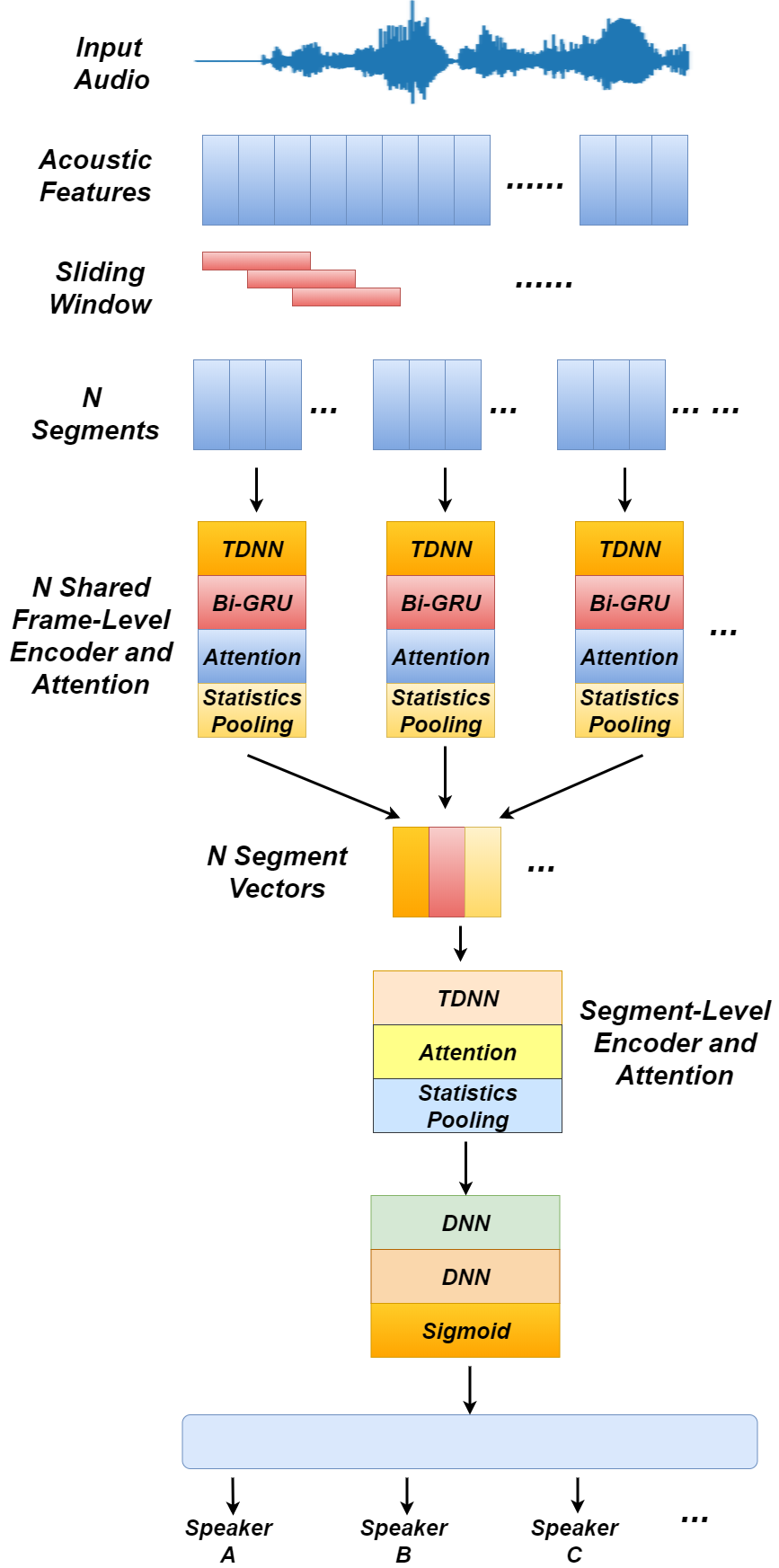}
\caption{Architecture of the proposed Hierarchical Attention Network.}
\label{proposed architecture}
\end{figure}

Figure \ref{proposed architecture} shows the architecture of the hierarchical attention network.
The network consists of several parts: a frame-level encoder and attention layer,
a segment-level encoder and attention layer, and two fully connected layers as a classifier.
Given the input acoustic frame vectors, the proposed model applies attention mechanism locally and globally. It predicts multiple speakers in the input utterance. 
The details of each part will be introduced in the following subsections. 

\vspace*{-3mm}
\subsection{Frame-Level Encoder and Attention}\label{Frame-Level Encoder and Attention}
\vspace*{-2mm}

An utterance is divided into $N$ segments: $\textbf{S} = \{\textbf{S}_1, \textbf{S}_2, \cdots, \textbf{S}_N\}$
using a sliding window with length $M$ and step $H$ . Each segment $\textbf{S}_i \in \mathcal {R}^{M \times L} = \{\textbf{x}_{i1}, \textbf{x}_{i,2}, \cdots, \textbf{x}_{i,M}\}$ contains $M$ $L$-dimensional acoustic frame vectors $\boldsymbol x_{i,t} \in \mathcal {R}^{1 \times L}$, 
where $i$ denotes the $i$th segment, $t$ denotes the $t$th frame,  $i \in \{1,\cdots N\}, t \in \{1, \dots, M\}$.

In the frame-level encoder, a TDNN \cite{peddinti2015time} is used on each segment, and followed by a 
bidirectional GRU \cite{chung2014empirical} in order to get information
from both directions of acoustic frames and contextual information.

The output of a frame-level encoder $\boldsymbol h_{i}=[\overrightarrow{\boldsymbol h}_{i}, \overleftarrow{\boldsymbol h}_{i}] \in 
\mathcal {R}^{M \times E} = \{\boldsymbol h_{i,1}, \boldsymbol h_{i,2}, \cdots, \boldsymbol h_{i,M}\}$
contains the information of the segment $\boldsymbol S_{i}$.

In the frame-level attention layer, a two-layer MLP is first used
to convert $\boldsymbol h_{i}$ into score vector $\boldsymbol z_{i}$, by which a
normalized importance weight vector $\boldsymbol \alpha_{i}$ can be computed via a softmax function \cite{yang2016hierarchical,rimer2004softprop}. 

\begin{equation}
\alpha_{i,t} = \frac{\mathrm{exp}(z_{i,t})}{\sum_{t=0}^{M} \mathrm{exp}(z_{i,t})}
\end{equation}

\begin{equation}\label{att}
z_{i,t} = \mathrm{Relu}(\boldsymbol h_{i,t} \boldsymbol W_{i,0}+\boldsymbol b_{i,0})\boldsymbol W_{i,1} ~~~~~~,
\end{equation}
where $z_{i,t}$ and $\alpha_{i,t}$ are a scalar score and normalized score for each time step $t$ respectively. $\boldsymbol W_{i,0} \in \mathcal {R}^{E \times E}$, $\boldsymbol b_{i,0} \in \mathcal {R}^{1 \times E}$ and $\boldsymbol W_{i,1} \in \mathcal {R}^{E \times 1}$ are the parameters of a two-layer MLP.
These parameters are shared when processing $N$ segments.
A weighted output of the frame-level encoder is computed by 
\begin{equation}
\boldsymbol A_{i,t} =  \alpha_{i,t} \boldsymbol h_{i,t}
\end{equation}
Following \cite{snyder2018x}, statistics pooling is applied on $\boldsymbol A_{i}$ to compute
its mean vector ($\boldsymbol \mu_{i}$) and std ($\boldsymbol \sigma_{i}$) 
vector over time.
A segment vector $\boldsymbol V_{S_{i}}$ is then obtained by concatenating the two vectors:
\begin{equation}
\boldsymbol V_{S_{i}} =  \mathrm{concatenate}(\boldsymbol \mu_{i}, \boldsymbol \sigma_{i})
\end{equation}

\vspace*{-2mm}
\subsection{Segment Level Encoder and Attention}
\vspace*{-2mm}
For the segment-level encoder and attention, the segment vector sequence is input to a stack of TDNN layers followed by a attention that descript in section \ref{Frame-Level Encoder and Attention}. 
as the omission of the GRU layer can well accelerate training when processing
a large number of samples.

The output of the frame level encoder and attention is $\boldsymbol V_{S} \in \mathcal {R}^{N \times E} = \{\boldsymbol V_{S_{1}}, \boldsymbol V_{S_{2}}, \cdots, \boldsymbol V_{S_{N}}\}$. The weight vector $\boldsymbol \alpha^s \in \mathcal {R}^{N \times 1} = \{\alpha^{s}_{1}, \alpha^{s}_{2}, \cdots, \alpha^{s}_{N}\}$ of segment level attention
can be computed as follows \cite{Pan2019AutomaticHA}:
\begin{equation}\label{att}
\begin{aligned}
\alpha^{s}_{i} &= \frac{\mathrm{exp}(z^{s}_{i})}{\sum_{i=0}^{N} \mathrm{exp}(z^{s}_{i})}\\
z^{s}_{i} &= \mathrm{Relu}(\boldsymbol V_{S_{i}} \boldsymbol W_{n,0}+\boldsymbol b_{n,0})\boldsymbol W_{n,1}~~~~~~,
\end{aligned}
\end{equation}
where $z^{s}_{i}$ and $\alpha^{s}_{i}$ are a scalar score and normalized score for each segment vector $\boldsymbol V_{S_{i}}$ respectively. $\boldsymbol W_{n,0} \in \mathcal {R}^{E \times E}$, $\boldsymbol b_{n,0} \in \mathcal {R}^{1 \times E}$ and $\boldsymbol W_{n,1} \in \mathcal {R}^{E \times 1}$ are the parameters of a two-layer MLP.
A vector is generated using a statistics pooling over all weighted segments:
\begin{equation}
\begin{aligned}
   \boldsymbol \mu_{U} =  mean(\sum_i \alpha^s_i \boldsymbol S_i)\\
   \boldsymbol \sigma_{U} = std(\sum_i \alpha^s_i \boldsymbol S_i)\\
   \boldsymbol V_U = \mathrm{concatenate}(\boldsymbol \mu_{U}, \boldsymbol \sigma_{U})   
\end{aligned}
\end{equation}

The final speaker identity classifier is constructed using a two-layer MLP followed by a sigmoid activation function \cite{ito1991representation} with $\boldsymbol V_U$ being its input. The final speaker identities are the output vector which contains the scores (between 1 and 0) for each speaker. The model is trained using binary cross entropy loss \cite{xu2017unsupervised}: 
\begin{equation}
\footnotesize
E_{bce} = - \sum_{n=1}^{N}||\boldsymbol {Y_{n}} \log \boldsymbol {\hat{Y_{n}}} + (1- \boldsymbol {Y_{n}}) \log (1-\boldsymbol {\hat{Y_{n}}})||
\end{equation}
, where $\boldsymbol {\hat{Y_{n}}}$ denotes the predicted score vector and $\boldsymbol Y_{n}$ denote the reference label vector, $N$ denotes the batch size.

\vspace*{-2mm}
\section{Experiments}\label{Experiments}
\vspace*{-2mm}
\subsection{Data}
\vspace*{-2mm}
\begin{figure}[t]
\centering
\includegraphics[height=4.5cm,width=7.5cm]{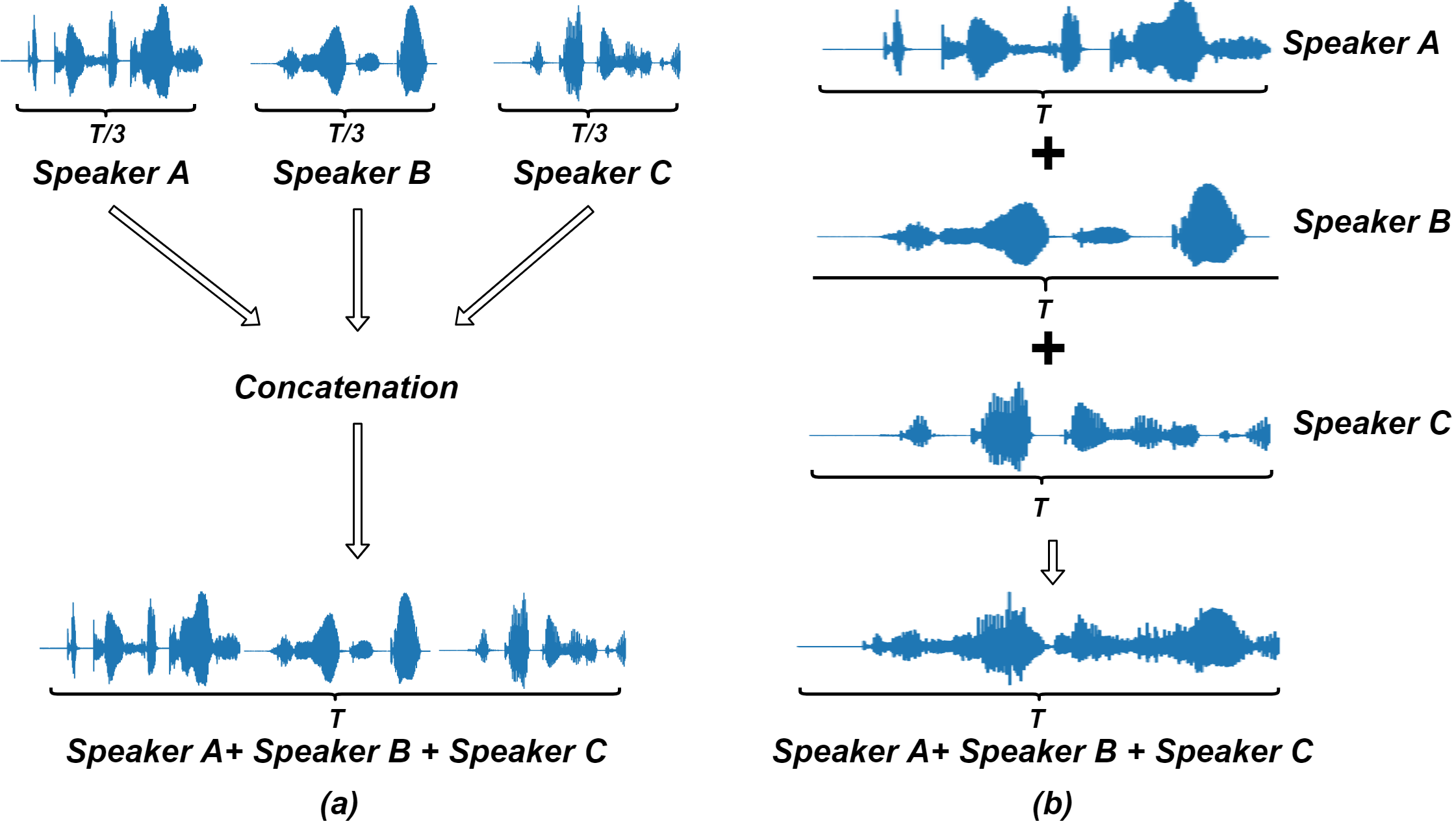}
\caption{The illustration of the data construction process. (a): Concat; (b): Overlap. }
\label{generate_data}
\end{figure}

\begin{table*}[h]
\renewcommand\arraystretch{1.0}
\centering
\begin{tabular}{cccccc}
    \hline
    Name & Original Dataset & Type & \#Select Speaker & \#Utterance Train & \#Utterance Test \\
    \hline
    SWBC-S & SWBC & Telephone & 254 & 6000 & 20,000 \\
    \hline    
    SWBC-L&SWBC& Telephone &  254 & 100,000 & 20,000\\
    \hline
    Vox-S & Voxceleb1 & Interview & 1000 & 15000 & 30,000\\
    \hline
    Vox-L & Voxceleb1 & Interview & 1000 & 150,000 & 30,000\\
    \hline

\end{tabular}
\caption{Details for the construction of the four datasets: SWBC-S, Vox-L, SWBC-S and Vox-L.}  
\label{data}
\end{table*}
In this work, Switchboard Cellular Part 1 (SWBC) \cite{swb} and Voxceleb1 \cite{nagrani2017voxceleb} dataset are used, as both of them are benchmark datasets and have been widely used in speaker identification. 
The SWBC dataset contains 130 hours telephone speech with 254 speakers 
(129 male and 125 female) under various environment conditions. The Voxceleb1 dataset contains 1251 speakers with more than 150,000 utterances collected in the wild. 20-dimensional MFCC \cite{tiwari2010mfcc} is used as the input acoustic features.

\vspace*{-2mm}
\subsubsection{Data Construction}
\vspace*{-2mm}
As there is no ready-made data for our task, new datasets are conducted manually by using the utterances from the Voxceleb1 and the SWBC dataset. To conduct weakly supervised training, two scenarios are designed: Overlap and Concat. Figure \ref{generate_data} (a) shows an example of the Concat scenario where the three speakers' voices are concatenated without an overlap. Figure \ref{generate_data} (b), shows an example of Overlap scenario where the three speakers' voices are completely overlapped. 

Based on the two scenarios above, in order to test the robustness of the proposed approach, for each of the two scenarios, 
four datasets are generated based on SWBC and Voxceleb1. Table \ref{data} shows the details of the four datasets. For the first dataset (SWBC-S, ``S'' represents small), SWBC dataset is used and each speaker occurs 30 times in the training set averagely. "SWBC-L" (``L'' represents large) contains more training data, each speaker occurs 200 times in the training data averagely, while the amount of the test data keeps the same. The small and large version of the datasets are used to test the robustness of the proposed model in small and large training data. Similar to the configurations in the SWBC based datasets, the datasets that based on Voxceleb1 also have small and large scenarios. 
In "Vox-S", 1000 speakers are randomly selected from the Voxceleb1 dataset. Each speaker occurs 30 times in the training set. In "Vox-L" dataset, each speaker occurs 300 times in the training set, while the test set is the same as "Vox-S".  For each of the eight datasets, the number of speakers in each utterance is randomly chosen from one to three in all of the datasets.

\vspace*{-2mm}
\subsection{Experiment Setup}
\vspace*{-2mm}
The proposed model is compared with two baselines: X-vectors \cite{snyder2018x} and Attentive X-vector (Att-Xvector) \cite{zhu2018self,okabe2018attentive,wang2018attention, rahman2018attention}. X-vectors contains TDNN based frame-level feature extractor, statistics pooling and DNN based segment-level feature extractor. Att-Xvectors uses an global attention mechanism after the TDNN based frame-level feature extractor. The proposed approach is denoted as "H-vector" and it is split into to scenarios: H-vector+sliding window and H-vector+static window. In H-vector+sliding window, the window length $M$ is set to 20 frames, and the step length $H$ is set to 10 frames. In H-vector+static window, the $M$ is set to 20 frames, and the $H$ is set to the same as $M$, which means there is no overlap for each local segments.  

In table \ref{data}, each of the four datasets contains two scenarios (Concat and Overlap). 
In the training process, for all of the eight datasets, the number of speakers in the generated utterances is not fixed, changing from one to three.
When the number of speakers is one, the generated utterance is the same as the original utterance. When the number of speakers are two or three, the output utterance contains multiple speakers with or without overlap. 

There are no overlaps between the training and test data. 
The length of all of the generated utterances are fixed at five seconds.

\vspace*{-2mm}
\subsection{Evaluation Metric}
\vspace*{-2mm}
In this work, equal error rate (EER) \cite{cheng2004method,murphy2012machine} is used as the evaluation metric, as it is widely used in multi-label audio tagging \cite{xu2017unsupervised}. The EER is defined as the point when the false negative (FN) equals to the false positive rate (FP) rate. EER is computed for each individual input and averaged across the whole test set \cite{cheng2004method}. 

\vspace*{-2mm}
\subsection{Implementation}
\vspace*{-2mm}
\begin{table}[h]
\renewcommand\arraystretch{1.0}
\setlength{\tabcolsep}{1.8mm}
\centering  
\footnotesize
\begin{tabular}{c|c|c|c}
\hline
Level& Model & Input & Output  \\
\hline
\multirow{4}{*}{Frame-Level}&
TDNN & (M,20) & (M,256)\\
&Bi-GRU &(M,256)& (M,512)\\
&Attention & (M,512)& (M,512)\\
&Statistics Pooling & (M,512)& (1,1024)\\
\hline

\multirow{5}{*}{Segment-Level}&
TDNN1 & (N,1024) & (N,512)\\
&TDNN2 & (N,512) & (N,512)\\
&TDNN3 & (N,512) & (N,1500)\\
&Attention & (N,1500)& (N,1500)\\
&Statistics Pooling & (N,1500)& (1,3000)\\
\hline

\multirow{2}{*}{Utterance-Level}&
DNN (512) & (1,3000) & (1,512)\\
&DNN (K)& (1,512)& (1,K)\\
\hline

\end{tabular}
\caption{Architecture of the proposed hierarchical attention network architecture, where K denotes the total number of speakers.}\label{model_summary}
\label{model_sum}
\end{table}

Table \ref{model_sum} shows the details of the proposed model architecture. The TDNN in both frame-level and segment-level encoder operates at the current time step.
Batch normalizations \cite{ioffe2015batch} are added after each layer except for attention layer. Adam optimiser \cite{Kingma2014AdamAM} is used for all experiments with $\beta_1=0.95$, $\beta_2=0.999$, and $\epsilon= 10^{-8}$. The initial learning rate is $10^{-4}$.

\begin{figure*}[h]
	\centering
	\includegraphics[height=5.5cm,width=17cm]{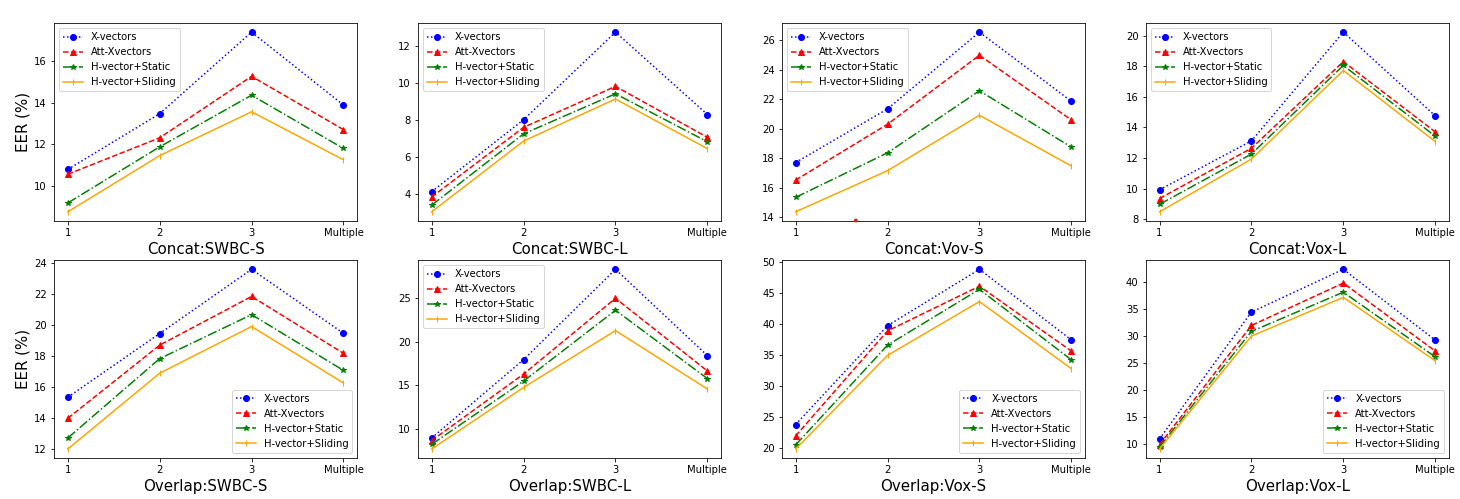}
	\caption{The results obtained using the four models (X-vectors, Attentive X-vectors, H-vector with static window and H-vector with sliding window) in different test conditions (1, 2, 3 or multiple speakers) on the eight designed datasets (SWBC-S, SWBC-L, Vox-S and Vox-L) and scenarios (Concat and Overlap). For all of the figures, the x-axis represents the number of speakers in test utterance. In H-vector with static window, the window size $M$ is 20 frames. In H-vector with sliding window, the window size $M$ is 20 frames, the step size $H$ is 10 frames.}
	\label{all_results}
\end{figure*}

\vspace*{-2mm}
\section{Results}\label{Results}
\vspace*{-3mm}

\begin{table}[h]
\renewcommand{\multirowsetup}{\centering}  
\renewcommand\arraystretch{1.0}
\setlength{\tabcolsep}{1mm}
\centering  
\footnotesize
\begin{tabular}{c|c|c|c|c|c}
\hline
\multirow{2}{*}{\textbf{Data Type}}& \multirow{2}{*}{\textbf{Window Size}} &\multicolumn{3}{c}{\textbf{EER (\%)}}\\

\cline{3-6}
& & SWBC-S& SWBC-L&Vox-S&Vox-L\\

\hline
\multirow{5}{*}{\textbf{Concat}}&
10 &  12.56&7.15 &18.29 & 13.69\\
&15&  11.87& 6.85& 18.08& 13.34\\
&20& \textbf{11.27}& \textbf{6.47}&\textbf{17.48}&\textbf{13.08}\\
&25&  11.69& 6.59& 17.81&13.29\\
&30&  12.11& 6.92& 18.21&13.66\\
\hline

\multirow{5}{*}{\textbf{Overlap}}&
10 & 17.81& 15.71& 34.37&26.46\\
&15& 16.89& 15.05& 33.48&25.85\\
&20& 16.24&\textbf{14.56} &32.77&\textbf{25.39}\\
&25& \textbf{15.99}&15.58 &\textbf{32.26} & 25.94\\
&30&16.59 &16.02 &32.86 &26.17\\
\hline
\end{tabular}
\caption{The obtained results of the proposed H-vector architecture using different window size $M$ (from 10 to 30 frames), step size $H$ is kept at 10 frames. }
\label{window_results}
\end{table}

\begin{table}[h]
\renewcommand{\multirowsetup}{\centering}  
\renewcommand\arraystretch{1.0}
\setlength{\tabcolsep}{1mm}
\centering  
\footnotesize
\begin{tabular}{c|c|c|c|c|c}
\hline
\multirow{2}{*}{\textbf{Data Type}}& \multirow{2}{*}{\textbf{Step Size}} &\multicolumn{3}{c}{\textbf{EER (\%)}}\\

\cline{3-6}
& & SWBC-S& SWBC-L&Vox-S&Vox-L\\

\hline
\multirow{5}{*}{\textbf{Concat}}&
5 &11.95 & 6.74 &18.01 &13.65\\
&10& \textbf{11.27}& 6.47& \textbf{17.48}&13.08\\
&15& 11.34& \textbf{6.29}& 17.98&\textbf{12.82}\\
&20& 11.45& 6.96& 18.21&13.15\\
&25& 11.86& 6.84& 18.56&13.42\\
\hline

\multirow{5}{*}{\textbf{Overlap}}&
5 & 16.49& 14.92& 33.87 &25.51\\
&10& \textbf{16.24}& 14.56& \textbf{32.77} &25.39\\
&15& 16.88& \textbf{14.13}&33.53&\textbf{24.86}\\
&20& 17.22& 14.82& 33.92&25.46\\
&25& 17.78& 15.11&34.25 &25.81\\
\hline
\end{tabular}
\caption{The obtained results of the proposed H-vector architecture using different step size $H$ (from 5 to 25 frames), window size $M$ is kept at 20 frames. }
\label{hop_results}
\end{table}

Figure \ref{all_results} shows the results obtained using the four models (X-vectors, Attentive X-vectors, H-vector with static window and H-vector with sliding window) in different test conditions (1, 2, 3 or multiple speakers) on the eight designed datasets (SWBC-S, SWBC-L, Vox-S and Vox-L) and scenarios (Concat and Overlap). In each figure, the X-axis represents the number of speakers in an utterance. ``''One, ``two'', ``three'' means the case where an utterance contains only one, two or three speakers, respectively. ``Multiple speaker'' means the combination of the three cases.

H-vector+Sliding window performs better in almost all of the conditions. The H-vector+static window performs better than the two baselines. These results show that capturing local and global information in weakly supervised speaker identification is helpful. The obtained results by X-vector is worst, this might because it treat each frame has equal importance. Comparing with Att-xvector, one of the reason of the improvement of the proposed H-vector might because of the distributed attention mechanism. Att-Xvector only applied attention mechanism globally.

Among all of the test conditions, the best results are obtained when the number of speakers in each utterance is one, and the worse case is when each utterance contains three speakers. This might due to the difficulty of the test conditions. A similar reason also occurs in the two different data construction scenarios (Concat and Overlap). In these two scenarios, the results obtained on Concat scenario is better than that on Overlap scenario. This might because when the speakers' voice are overlapped together, it is more difficult to distinguish different speakers. However, the proposed H-vector+sliding window performs better than the baselines in different test conditions and different data construction scenarios. 

Moreover, when the training data is small, the proposed H-vector+sliding window still performs better than the baselines and H-vector+static window, reaching 11.5 \% and 3.4 \% relative improvement than X-vectors and Att-Xvectors in SWBC-S dataset in Concat scenario. It shows robustness of the proposed H-vector+sliding window when there is no enough training data. 

In order to test the effectiveness of the window size (M) and step size (H), Table \ref{window_results} and \ref{hop_results} show the obtained results using the proposed H-vector+sliding window when using different window size and step size. In Overlap scenario, the equal error rate is more sensitive to the change of window size and step size. This might because in Overlap scenario, different speaker signals are overlapped in time domain, some speaker features might influence to each other. Different window size and step size allows the frame-level encoder and attention to capture more local features. Furthermore, in most of the cases, the best results is obtained when the window size is 20 frames, the step size is 10 frames, in which the step size is set to the half-size of the window size.

\vspace*{-3mm}
\section{Conclusion and Future Work}\label{Conclusion and Future Work}
\vspace*{-2mm}
In this work, a hierarchical attention network is proposed to solve the weakly labelled speaker identification problem. The input utterance is split into each local segments using a sliding window. Frame-level and segment-level encoder and attention capture speaker information locally and globally. The experiments are done with different test conditions and different amount of training data. The obtained results show that the proposed hierarchical attention network with sliding window performs better than X-vector and Attentive Xvector baselines, as well as hierarchical attention network with static window. 
In the future work, more complex network architectures and larger dataset such as Voxceleb2 will be investigated. 

\begin{center}
	\large{\textbf{Acknowledgement}}
\end{center}
\vspace*{-2mm}
This work was in part supported by Innovate UK Grant number 104264.

\newpage
\bibliographystyle{IEEEtran}

\bibliography{mybib}


\end{document}